\documentclass[twocolumn]{aastex631}
\usepackage{amsmath} 

\mathchardef\mhyphen="2D
\usepackage{CJKutf8}

\begin{document}

\begin{CJK*}{UTF8}{gbsn}
\title{Carving Out the Inner Edge of the Period Ratio Distribution through Giant Impacts}
\author[0009-0007-4983-9850]{Kaitlyn Chen}
\altaffiliation{Both these authors contributed equally to this manuscript and reserve the right to list themselves as first author on their respective CVs.}
\affiliation{Department of Physics, Harvey Mudd College, Claremont, CA 91711, USA}
\author[0009-0000-7595-5520]{Oswaldo Cardenas}
\altaffiliation{Both these authors contributed equally to this manuscript and reserve the right to list themselves as first author on their respective CVs.}
\affiliation{Department of Physics, Harvey Mudd College, Claremont, CA 91711, USA}
\author{Brandon Bonifacio}
\affiliation{Department of Physics, Harvey Mudd College, Claremont, CA 91711, USA}
\author{Nikolas Hall}
\affiliation{Department of Physics, Harvey Mudd College, Claremont, CA 91711, USA}
\author{Rori Kang}
\author[0000-0002-9908-8705]{Daniel Tamayo}
\altaffiliation{Corresponding author: \href{mailto:dtamayo@hmc.edu}{dtamayo@hmc.edu}}
\affiliation{Department of Physics, Harvey Mudd College, Claremont, CA 91711, USA}

\begin{abstract}

The distribution of orbital period ratios between adjacent observed exoplanets is approximately uniform, but exhibits a strong falloff toward close orbital separations.
We show that this falloff can be explained through past dynamical instabilities carving out the period ratio distribution.
Our suite of numerical experiments would have required $\sim 3$ million CPU-hours through direct N-body integrations, but was achieved with only $\approx 50$ CPU-hours by removing unstable configurations using the Stability of Planetary Orbital Configurations Klassifier (SPOCK) machine learning model.
This highlights the role of dynamical instabilities in shaping the observed exoplanet population, and shows that the inner part of the period ratio distribution provides a valuable observational anchor on the giant impact phase of planet formation.

\end{abstract}


\section{Introduction}
\end{CJK*}


Whether protoplanets grow primarily through the accretion of planetesimals \citep[e.g.,][]{Kokubo00} or through pebble accretion \citep[see review by][]{Johansen17}, planet formation theory typically converges on a final phase of giant impacts and gravitational scatterings that sets the final masses and orbital configurations of the exoplanets we observe today \citep[e.g.,][]{Goldreich04, Izidoro15, Izidoro17, Poon20, Goldberg22, Lammers23, Ghosh24}.
In this picture, the distribution of exoplanets' orbital elements should have been carved out and reshaped by dynamical instabilities \citep{Volk15, Pu15}.

\begin{figure}[!ht]
     \centering
    \includegraphics[width=0.45\textwidth]{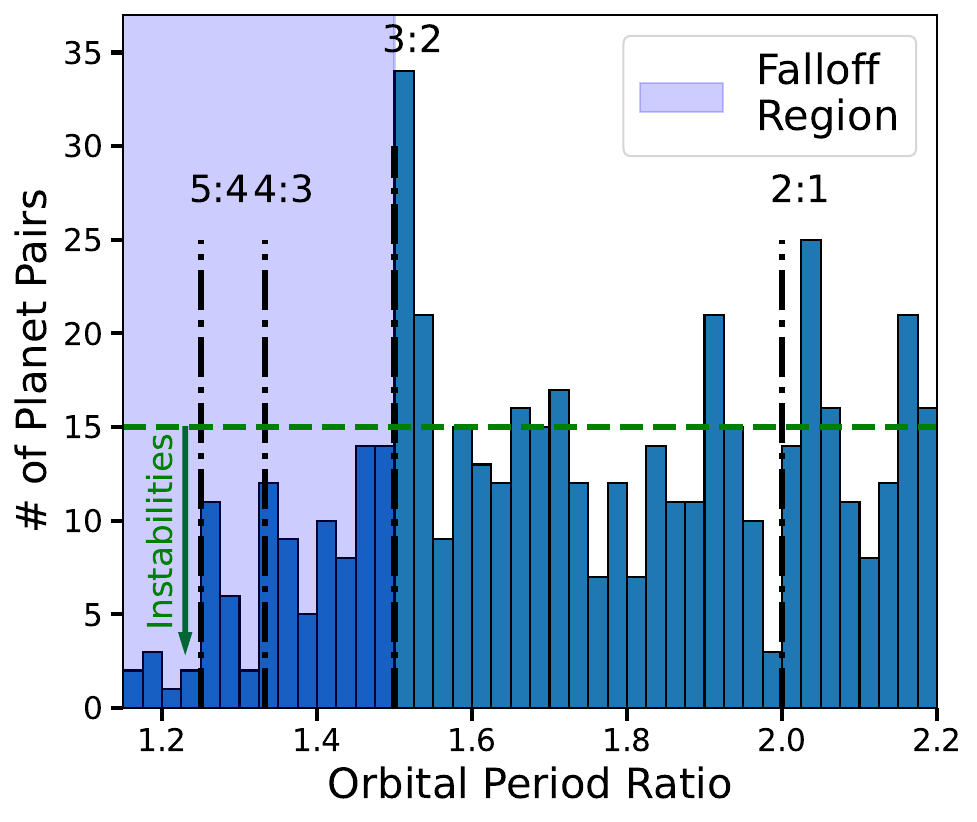}
     \caption{Histogram of the period ratio between adjacent exoplanets in the NASA exoplanet archive. Dotted vertical lines represent the location of first and second order MMRs. The horizontal dashed green line shows the approximately uniform distribution of period ratios beyond 1.5. In this paper we show that the falloff in the shaded blue region toward close separations can be explained through dynamical instabilities.}
     \label{Histroperiod}
    \end{figure}

The roughly uniform distribution of period ratios between adjacent planets approximated by the dashed green line in Fig.\:\ref{Histroperiod} between period ratios of 1.5 and 2.2 suggests a chaotic giant impact phase that randomizes orbital periods \citep{Fabrycky14}\footnote{The modest pileups near integer period ratios have been explained through a variety of mechanisms specific to such near-resonant configurations \citep{Lithwick12, Batygin12, Petrovich15, Chatterjee15, Wu24}. At larger period ratios beyond $\approx 2.2$, there is a falloff as shown in Fig. 5 of \cite{Weiss23}, but this is likely due to observational biases.}.
Notably, however, the distribution falls off toward period ratios approaching unity at the tightest orbital separations.
Several authors have investigated analytical and semi-analytical stability criteria beyond which planetary configurations should no longer be observed \citep{Wisdom80, Gladman93, Quillen11, deck13, Laskar17, Petit17, Petit18, Hadden18, Petit20, Tamayo21, Rath22, lammers2024}.
However, this boundary depends on many physical and orbital parameters, and remains imperfectly understood theoretically.
We therefore opt for a numerical approach.

Several groups have modeled the giant impact phase directly through computationally expensive N-body integrations \citep{Chambers98, Agnor99, Hansen12, Hansen13, Dawson16}.
An alternative approach proposed by \cite{Tremaine15} is to assume the limiting case where this phase is sufficiently chaotic that memory of the initial conditions is lost, and to approximate the phase space of possible orbital configurations as uniformly filled.
The resulting planetary population that we see today is then the subset of these outcomes that are dynamically stable over the system's lifetime.
Assuming more widely separated systems are stable over their Gyr lifetimes, this ``ergodic" hypothesis explains the approximately uniform distribution of period ratios beyond 1.5.
In this paper we evaluate the hypothesis that an initially uniform distribution of period ratios that previously extended below period ratios of 1.5 would have been carved out by dynamical instabilities to produce the observed falloff toward small separations.

We begin this paper by describing our initial conditions and methods for determining stability in Sec.\:\ref{sec:methods}. In Sec.\:\ref{Results Page}, we then compare our resulting synthetic populations against observed systems, and explore the robustness of our results as we vary our assumed population parameters. Finally, we summarize and discuss the implications of the results in Sec. \:\ref{Sec:Conclusions}. 

\section{Methods} \label{sec:methods}

Because dynamical stability depends on a number of factors in addition to the period ratio plotted in Fig.\:\ref{Histroperiod}, we begin by considering the initialization of our primordial synthetic population.
In our nominal model, we aim to draw masses and orbital parameters close to the observed exoplanet population as outlined below.
We later consider how varying some of our population parameters affects the validity of our results in Sec.\:\ref{sec:sensitivity}.

\subsection{Creating Initial Populations}

We consider five characteristics of exoplanet systems when constructing our simulations: mass, eccentricity, period ratio, correlation of period ratios, and number of exoplanets in the system. 

Perhaps most importantly, dynamical instabilities are qualitatively different for isolated pairs of planets vs. systems with three or more planets.
In two-planet systems, there exists a critical ``Hill-limit" orbital separation beyond which the pair is guaranteed to never undergo close encounters \citep{Marchal82, Gladman93}. 
This boundary separates pairs that are too close and typically destabilize within $\sim$ a hundred orbits, from pairs that are stable beyond systems' Gyr lifetimes.
By contrast, systems with three or more planets exhibit instabilities for significantly wider interplanetary separations, and the time to such instabilities spans the full dynamic range from orbital timescales to many billions of orbits \citep[e.g.,][]{Chambers96, Smith09, Obertas17}.
This is thought to be due to the effects of three-body MMRs \citep{Quillen11, Petit20, Rath22, lammers2024}, as well as long-term oscillations in the eccentricities that cause MMRs to adiabatically expand and contract, thereby sweeping out regions of chaos \citep{Tamayo21, Yang24}.
Given that, on average, observed systems have three planets with orbital periods within 400 days \citep{Zhu18} and such systems are the lowest multiplicity to exhibit the general dynamical behavior of compact multiplanet systems, we choose to generate synthetic populations of three-planet systems.

We draw our synthetic populations from the set of observed systems taken from the NASA exoplanet archive on August 4, 2024\footnote{Planetary Systems Composite Parameters Table, doi:10.26133/NEA13.}. 
For simplicity, we begin by filtering for all planets in 3+ planet systems that have an inner period ratio ($P_2/P_1$) below 1.5\footnote{We also performed tests with the population of 3+ planet systems where the outer pair had a period ratio $P_3/P_2$ below 1.5, and found a similar (slightly better) match between the synthetic and observed populations reported below.}. We then calculate planet-star mass ratios and remove any planets where these are undefined (due to either a missing planet, or stellar mass), or giant planets with mass ratios $>10^{-4} \approx 2$ Neptune masses. After filtering, we are left with 73 compact pairs.

Given that Newtonian gravity is scale invariant, we rescale all our synthetic systems to have central stars of $1 M_\odot$ and an inner planet with an orbital period of $1$ yr. We then draw the planet-star mass ratios and orbital eccentricity for each planet randomly and independently from the observed distributions described above (with replacement).

\begin{figure*} [!ht]
    \centering
    \includegraphics[width=16cm]{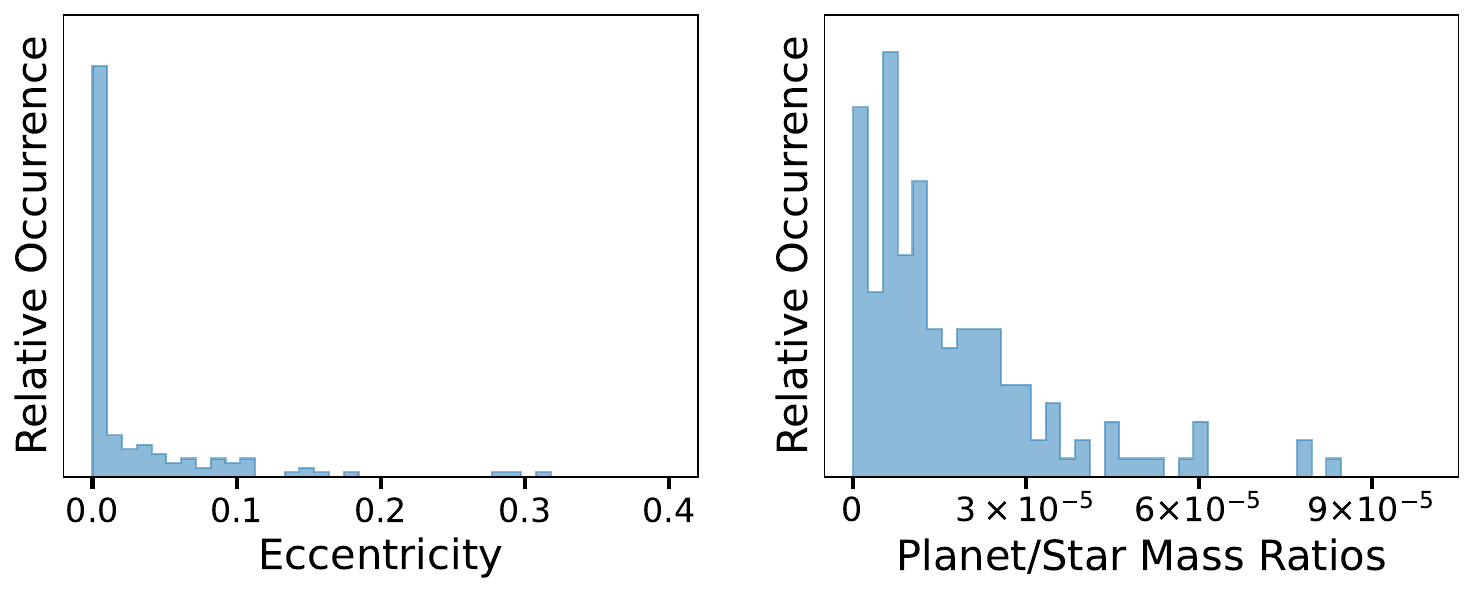}
    \caption{Distribution of filtered eccentricities used to generate synthetic planetary systems (left panel). Right panel shows the corresponding distribution of planet-star mass ratios.}%
    \label{LosDos}%
\end{figure*}

Graphs plotting both the observed and synthetic distributions for the eccentricities and mass ratios can be found in Figure \ref{LosDos}. 

We randomly assign the longitude of pericenter for each orbit uniformly between [0,2$\pi$). 
For simplicity, we also make all systems co-planar, since typical mutual inclinations of transiting exoplanets of less than a few degrees were found by \cite{Tamayo21} to have a negligible effect on dynamical stability on timescales of less than a billion orbits.

We choose the period ratio for the inner planet pair randomly between our range of interest of 1.1 to 1.5. For the outer pair, we try to capture the observed preference toward uniform orbital spacings \citep{Weiss18}, since this also should affect dynamical stability \citep{Petit20, Rath22}.
We do this by randomly drawing from a Gaussian distribution centered at the inner period ratio (corresponding to uniform spacing) with a standard deviation that we empirically determine in order to reproduce the observed spread.
We quantify this deviation from uniform spacings by calculating a normalized dispersion $D_i$ for each system
\begin{align}
D_i= \frac{\sigma_i}{\mu_i}.
\label{D_trio}
\end{align}
where $i$ is an index labeling the multiplanet system, and $\mu_i$ and $\sigma_i$ are the corresponding mean and standard deviation of the $\log_{10}$ of the period ratios between adjacent planets in that system. 
From the NASA Exoplanet Archive, we calculated a mean dispersion for our observed sample (filtered as described above) of $D=0.30$.

\subsection{\texttt{SPOCK} for Stability Determination}
Traditionally, one would simulate dynamical evolution via direct numerical integrations; however, this method is computationally expensive. Utilizing the WHFast integrator \citep{Rein2015} in the \texttt{REBOUND} N-body package \citep{Rein12}, a $10^9$-orbit integration takes $\approx 6$ CPU hours using a 2.1 GHz Intel Xeon Silver 4116. For the numerical experiments we perform in this paper, we estimate N-body methods would require almost $\approx 3$ million CPU hours.

We therefore instead use the Stability of Planetary Orbital Configurations Klassifier (\texttt{SPOCK}), an open-access machine learning model that predicts the stability of compact systems of three or more planets \citep{Tamayo16, Tamayo20}. \texttt{SPOCK} runs short $10^4$-orbit integrations, and calculates dynamically informed features from which the gradient-boosted decision-tree model is trained to predict stability over $10^9$ orbits. SPOCK is roughly $10^5$ times faster than direct integration, and outputs a probability of survival between 0 and 1.
This reduces the computational cost to $\approx 50$ CPU hours. 

\subsection{Predicting the Period Ratio Falloff}

For each experiment, we follow the procedure outlined in Section 2.1 to repeatedly generate synthetic 3-planet systems. 
We build up our simulated population by retaining each sampled system with a probability given by its corresponding likelihood of stability as estimated by SPOCK.
We stop once we reach a population that contains the same number of compact pairs as the observations (73). 
This allows us to naturally weight systems with higher probabilities of stability more heavily \citep{tamayo21b}.

\section{Results} \label{Results Page}

\begin{figure}[!ht]  \label{PeriodRatio}
 \centering
\includegraphics[width=0.4\textwidth]{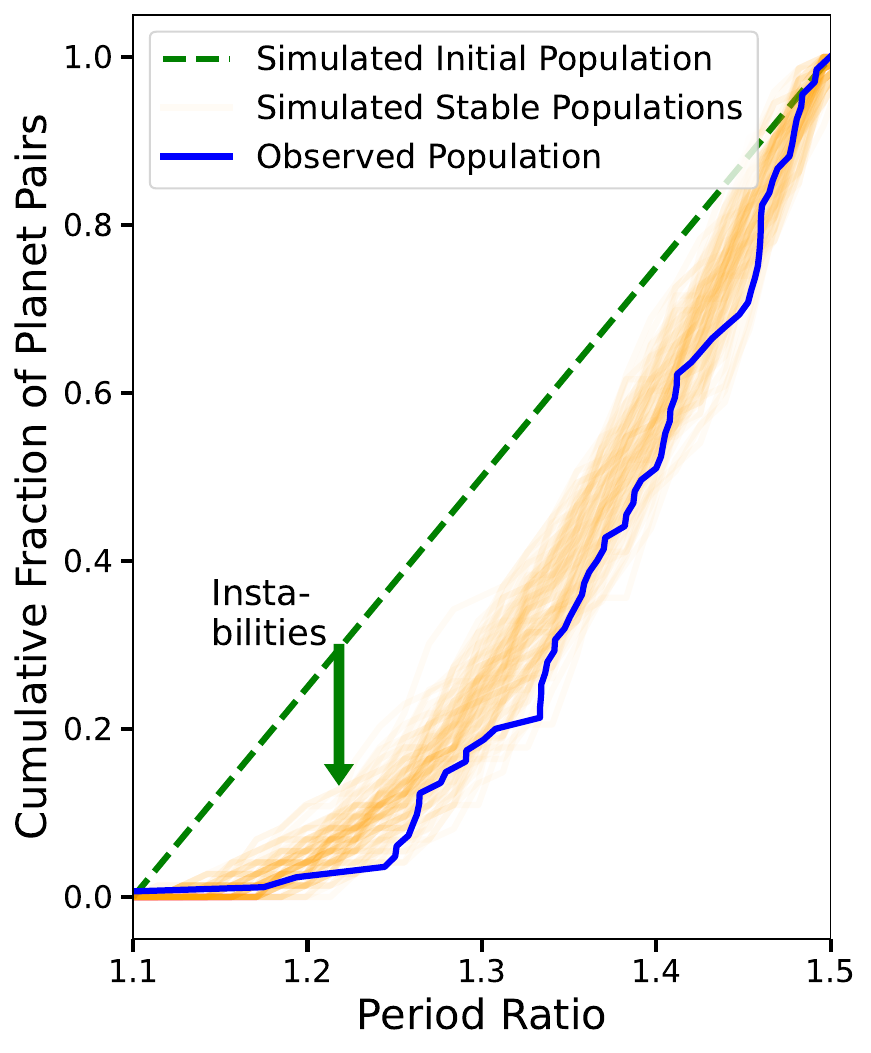}
 \caption{Cumulative distribution of orbital period ratios between adjacent planets in observed systems of three or more planets (blue). Each orange line represents a different realization drawing an initial  uniform synthetic population (green dashed line) and then filtering out unstable configurations. The average p-value estimating the probability a given orange curve is drawn from the same distribution as the observations (blue) is 0.48}
 
\end{figure}

We plot the cumulative distribution function (CDF) of surviving period ratios in one such trial as an orange line in Fig.\:\ref{PeriodRatio}, and perform a Kolmogorov–Smirnov test between the synthetic (orange) and observed (blue) CDFs yielding a p-value that the two curves were drawn from the same underlying distribution. 
We then repeat this procedure in 100 trials, obtaining 100 random realizations as orange lines in Fig.\:\ref{PeriodRatio}.

We observe an approximate correspondence with observations, with an average p-value of 48\%, consistent with having been drawn from the same distribution as the real systems.

\begin{figure}[!ht]
 \centering
\includegraphics[width=0.4\textwidth]{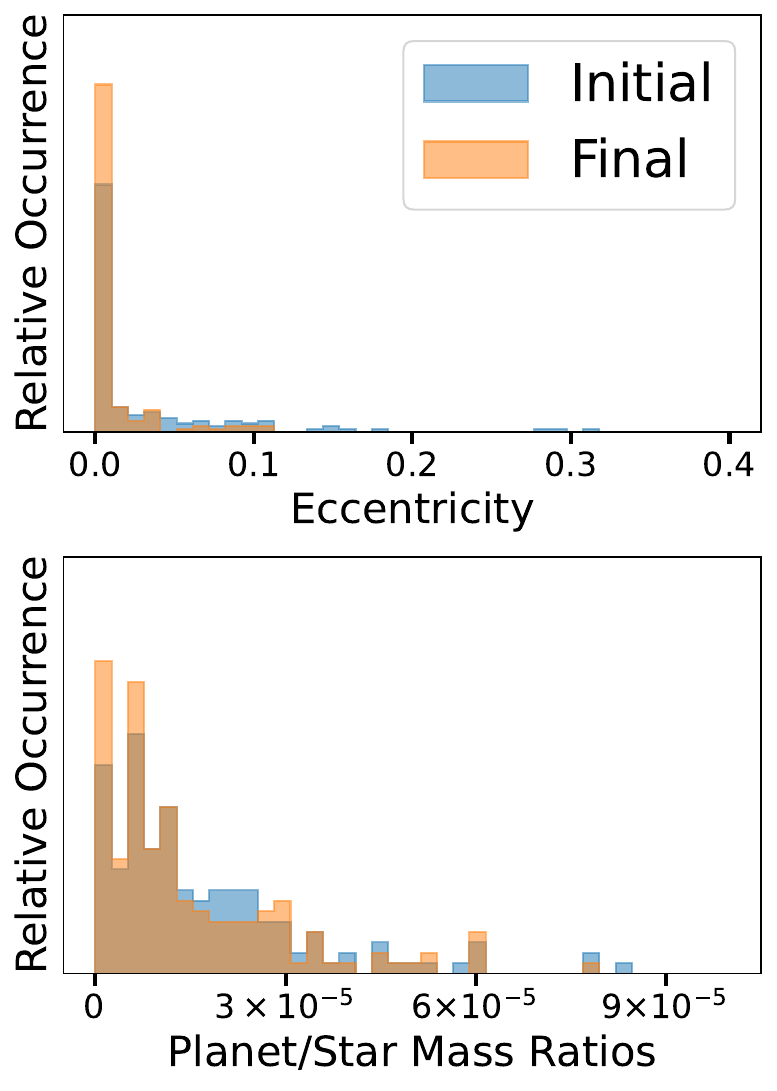}
 \caption{Initial distributions for the planet-star mass ratios and orbital eccentricities (orange) drawn from the observed exoplanet population. In blue we plot the resulting distributions after filtering for orbital stability as described in the text.
 \label{MassSPOCK}}
\end{figure}

In Figure \ref{MassSPOCK}, we compare our distribution of mass ratios and orbital eccentricities before (blue) and after (orange) filtering for stability.
As expected, there is a preference toward lower masses and orbital eccentricities.
We note that we draw the initial populations (blue) from the observed distribution of exoplanets.
In principle it should be the surviving population (in orange) that matches observations; however, given the small differences in Fig.\:\ref{MassSPOCK}, we do not try to correct for this effect.

\subsection{Sensitivity Tests} \label{sec:sensitivity}

We now assess the sensitivity of our results to our various population parameters, which are both uncertain and potentially observationally biased.

\begin{figure*}
\includegraphics[width=.99\textwidth]{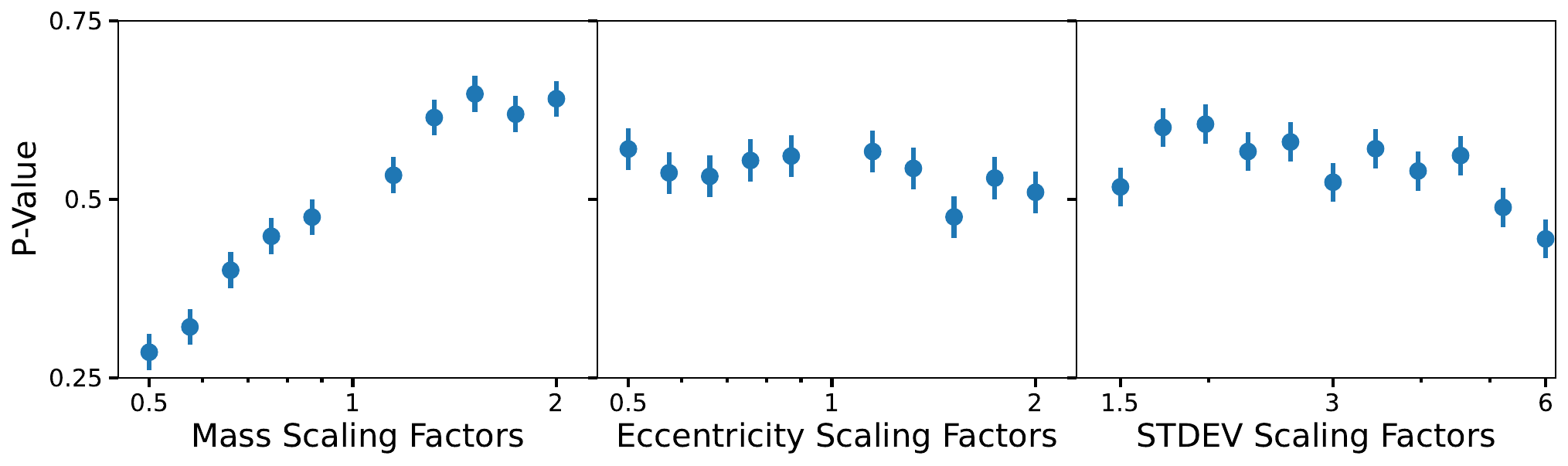}
 \caption{Each point in the three panels displays the mean p-value across 100 realizations, estimating the probability that the synthetic planet population matches the observed one (error bars denote one standard deviation). The corresponding population parameter (mass, eccentricity, normalized dispersion of interplanetary separations, Eq.\:\ref{D_trio}) is scaled by the value along the $x-axis$. The plots demonstrate the consistency of our simulated systems against observations over a broad range of population parameters.}
 \label{EccentrityPvalue}
\end{figure*}

We first perform a sequence of experiments, where we reproduce 100 synthetic trials like in Fig.\:\ref{PeriodRatio}, except after drawing the mass-ratio for each planet, we multiply all the masses by a common scaling factor.
We vary the scaling factor logarithmically from 0.5 to 2 in the left panel of Fig.  \ref{EccentrityPvalue}, where each blue point corresponds to the mean (and error bars to the standard deviation) of the p-values across the 100 trials of that experiment.

We see that we retain qualitatively similar results over the fairly broad range of parameters probed, showing that the result is robust against these choices. 
As a simple test of how the biases inherent to different observational methods affect our results, we also considered only the subset of planet pairs with masses measured through radial velocities 71, dropping the remainder (largely derived from mass-radius relations).
We find this yields an even larger p-value of 0.66.

The middle panel does the equivalent test, scaling the orbital eccentricities. 
We see that it does not significantly affect the p-values, mostly because a large fraction of our observed sample have circular orbits.
Of course, eccentricities of zero in the Exoplanet Archive do not necessarily imply circular orbits, but typically reflect that the eccentricities were too small to measure observationally.
As an opposite limiting case, if we instead use our nominal scaling factor of 1, but remove all zero-eccentricity planets, we obtain a p-value of 0.25. We interpret this as a lower bound, given that this sample should instead be biased toward high eccentricities (since high eccentricities are easier to measure).
Indeed, we find that the average non-zero eccentricity in this sample is $0.067$, somewhat higher than population-level constraints from transit durations in multi-planet systems that fit a Rayleigh distribution with a Rayleigh parameter of $\sigma_e = 0.049$ \citep{vanEylen15}.
 
We similarly find that varying the dispersion multiplier to vary the uniformity of our period ratios does not significantly affect our results.

\section{Discussion and Conclusions} \label{Sec:Conclusions}

In summary, we have used the SPOCK stability classifier \citep{Tamayo20} to show that an initially randomized (uniform) distribution of period ratios extending to close separations would undergo dynamical instabilities over $\sim 10-100$ Myr ($10^9$ orbits) and result in a surviving population that approximately matches the observed falloff of planet pairs toward compact spacings (Fig.\:\ref{PeriodRatio}).
We additionally find that this result is robust against modest changes in our assumed population parameters (Fig.\:\ref{EccentrityPvalue}).
The code for running the experiments and generating the associated figures in this paper is available at \url{ https://github.com/Waldoroni/InnerEdgePaper}.

This work highlights the role of dynamical instabilities in carving out and shaping the observed exoplanet sample \citep[e.g.,][]{Volk15, Pu15, Izidoro17}, and how tools for rapidly evaluating the stability of planetary configurations \citep[e.g,][]{Tamayo20} enable filtering synthetic populations of young systems for stability to compare against the mature observational sample.

\begin{acknowledgements}
We are grateful to an anonymous reviewer whose insightful comments improved the quality and clarity of this manuscript.
This work was funded in part by The Rose Hills Foundation Science and Engineering Summer Undergraduate Research Fellowship program. 
The presented numerical calculations were made possible by computational resources provided through an endowment by the Albrecht family.
This research has made use of the NASA Exoplanet Archive, which is operated by the California Institute of Technology, under contract with the National Aeronautics and Space Administration under the Exoplanet Exploration Program.
\end{acknowledgements}

\bibliography{Bib}

\begin{thebibliography}{}
\expandafter\ifx\csname natexlab\endcsname\relax\def\natexlab#1{#1}\fi
\providecommand{\url}[1]{\href{#1}{#1}}
\providecommand{\dodoi}[1]{doi:~\href{http://doi.org/#1}{\nolinkurl{#1}}}
\providecommand{\doeprint}[1]{\href{http://ascl.net/#1}{\nolinkurl{http://ascl.net/#1}}}
\providecommand{\doarXiv}[1]{\href{https://arxiv.org/abs/#1}{\nolinkurl{https://arxiv.org/abs/#1}}}

\bibitem[{Agnor {et~al.}(1999)Agnor, Canup, \& Levison}]{Agnor99}
Agnor, C.~B., Canup, R.~M., \& Levison, H.~F. 1999, Icarus, 142, 219

\bibitem[{Batygin \& Morbidelli(2012)}]{Batygin12}
Batygin, K., \& Morbidelli, A. 2012, The Astronomical Journal, 145, 1

\bibitem[{Chambers \& Wetherill(1998)}]{Chambers98}
Chambers, J., \& Wetherill, G. 1998, Icarus, 136, 304

\bibitem[{Chambers {et~al.}(1996)Chambers, Wetherill, \& Boss}]{Chambers96}
Chambers, J., Wetherill, G., \& Boss, A. 1996, Icarus, 119, 261

\bibitem[{Chatterjee \& Ford(2015)}]{Chatterjee15}
Chatterjee, S., \& Ford, E.~B. 2015, The Astrophysical Journal, 803, 33

\bibitem[{Dawson {et~al.}(2016)Dawson, Lee, \& Chiang}]{Dawson16}
Dawson, R.~I., Lee, E.~J., \& Chiang, E. 2016, The Astrophysical Journal, 822, 54

\bibitem[{{Deck} {et~al.}(2013){Deck}, {Payne}, \& {Holman}}]{deck13}
{Deck}, K.~M., {Payne}, M., \& {Holman}, M.~J. 2013, \apj, 774, 129, \dodoi{10.1088/0004-637X/774/2/129}

\bibitem[{Fabrycky {et~al.}(2014)Fabrycky, Lissauer, Ragozzine, Rowe, Steffen, Agol, Barclay, Batalha, Borucki, Ciardi, {et~al.}}]{Fabrycky14}
Fabrycky, D.~C., Lissauer, J.~J., Ragozzine, D., {et~al.} 2014, The Astrophysical Journal, 790, 146

\bibitem[{{Ghosh} \& {Chatterjee}(2024)}]{Ghosh24}
{Ghosh}, T., \& {Chatterjee}, S. 2024, \mnras, 527, 79, \dodoi{10.1093/mnras/stad2962}

\bibitem[{Gladman(1993)}]{Gladman93}
Gladman, B. 1993, Icarus, 106, 247

\bibitem[{{Goldberg} \& {Batygin}(2022)}]{Goldberg22}
{Goldberg}, M., \& {Batygin}, K. 2022, \aj, 163, 201, \dodoi{10.3847/1538-3881/ac5961}

\bibitem[{Goldreich {et~al.}(2004)Goldreich, Lithwick, \& Sari}]{Goldreich04}
Goldreich, P., Lithwick, Y., \& Sari, R. 2004, Annu. Rev. Astron. Astrophys., 42, 549

\bibitem[{Hadden \& Lithwick(2018)}]{Hadden18}
Hadden, S., \& Lithwick, Y. 2018, The Astronomical Journal, 156, 95

\bibitem[{Hansen \& Murray(2012)}]{Hansen12}
Hansen, B.~M., \& Murray, N. 2012, The Astrophysical Journal, 751, 158

\bibitem[{Hansen \& Murray(2013)}]{Hansen13}
---. 2013, The Astrophysical Journal, 775, 53

\bibitem[{Izidoro {et~al.}(2015)Izidoro, Morbidelli, Raymond, Hersant, \& Pierens}]{Izidoro15}
Izidoro, A., Morbidelli, A., Raymond, S.~N., Hersant, F., \& Pierens, A. 2015, Astronomy \& Astrophysics, 582, A99

\bibitem[{Izidoro {et~al.}(2017)Izidoro, Ogihara, Raymond, Morbidelli, Pierens, Bitsch, Cossou, \& Hersant}]{Izidoro17}
Izidoro, A., Ogihara, M., Raymond, S.~N., {et~al.} 2017, Monthly Notices of the Royal Astronomical Society, 470, 1750

\bibitem[{Johansen \& Lambrechts(2017)}]{Johansen17}
Johansen, A., \& Lambrechts, M. 2017, Annual Review of Earth and Planetary Sciences, 45, 359

\bibitem[{Kokubo \& Ida(2000)}]{Kokubo00}
Kokubo, E., \& Ida, S. 2000, Icarus, 143, 15

\bibitem[{{Lammers} {et~al.}(2023){Lammers}, {Hadden}, \& {Murray}}]{Lammers23}
{Lammers}, C., {Hadden}, S., \& {Murray}, N. 2023, \mnras, 525, L66, \dodoi{10.1093/mnrasl/slad092}

\bibitem[{Lammers {et~al.}(2024)Lammers, Hadden, \& Murray}]{lammers2024}
Lammers, C., Hadden, S., \& Murray, N. 2024, The instability mechanism of compact multiplanet systems.
\newblock \doarXiv{2403.17928}

\bibitem[{Laskar \& Petit(2017)}]{Laskar17}
Laskar, J., \& Petit, A. 2017, Astronomy \& Astrophysics, 605, A72

\bibitem[{Lithwick \& Wu(2012)}]{Lithwick12}
Lithwick, Y., \& Wu, Y. 2012, The Astrophysical Journal Letters, 756, L11

\bibitem[{Marchal \& Bozis(1982)}]{Marchal82}
Marchal, C., \& Bozis, G. 1982, Celestial Mechanics, 26, 311

\bibitem[{Obertas {et~al.}(2017)Obertas, Van~Laerhoven, \& Tamayo}]{Obertas17}
Obertas, A., Van~Laerhoven, C., \& Tamayo, D. 2017, Icarus, 293, 52

\bibitem[{Petit {et~al.}(2017)Petit, Laskar, \& Bou{\'e}}]{Petit17}
Petit, A.~C., Laskar, J., \& Bou{\'e}, G. 2017, Astronomy \& Astrophysics, 607, A35

\bibitem[{Petit {et~al.}(2018)Petit, Laskar, \& Bou{\'e}}]{Petit18}
---. 2018, Astronomy \& Astrophysics, 617, A93

\bibitem[{Petit {et~al.}(2020)Petit, Pichierri, Davies, \& Johansen}]{Petit20}
Petit, A.~C., Pichierri, G., Davies, M.~B., \& Johansen, A. 2020, Astronomy \& Astrophysics, 641, A176

\bibitem[{Petrovich {et~al.}(2013)Petrovich, Malhotra, \& Tremaine}]{Petrovich15}
Petrovich, C., Malhotra, R., \& Tremaine, S. 2013, The Astrophysical Journal, 770, 24

\bibitem[{{Poon} {et~al.}(2020){Poon}, {Nelson}, {Jacobson}, \& {Morbidelli}}]{Poon20}
{Poon}, S. T.~S., {Nelson}, R.~P., {Jacobson}, S.~A., \& {Morbidelli}, A. 2020, \mnras, 491, 5595, \dodoi{10.1093/mnras/stz3296}

\bibitem[{Pu \& Wu(2015)}]{Pu15}
Pu, B., \& Wu, Y. 2015, The Astrophysical Journal, 807, 44

\bibitem[{Quillen(2011)}]{Quillen11}
Quillen, A.~C. 2011, Monthly Notices of the Royal Astronomical Society, 418, 1043

\bibitem[{Rath {et~al.}(2022)Rath, Hadden, \& Lithwick}]{Rath22}
Rath, J., Hadden, S., \& Lithwick, Y. 2022, The Astrophysical Journal, 932, 61

\bibitem[{Rein \& Liu(2012)}]{Rein12}
Rein, H., \& Liu, S.-F. 2012, Astronomy \& Astrophysics, 537, A128

\bibitem[{Rein \& Tamayo(2015)}]{Rein2015}
Rein, H., \& Tamayo, D. 2015, Monthly Notices of the Royal Astronomical Society, 452, 376

\bibitem[{Smith \& Lissauer(2009)}]{Smith09}
Smith, A.~W., \& Lissauer, J.~J. 2009, Icarus, 201, 381

\bibitem[{Tamayo {et~al.}(2021)Tamayo, Gilbertson, \& Foreman-Mackey}]{tamayo21b}
Tamayo, D., Gilbertson, C., \& Foreman-Mackey, D. 2021, Monthly Notices of the Royal Astronomical Society, 501, 4798

\bibitem[{{Tamayo} {et~al.}(2021){Tamayo}, {Murray}, {Tremaine}, \& {Winn}}]{Tamayo21}
{Tamayo}, D., {Murray}, N., {Tremaine}, S., \& {Winn}, J. 2021, \aj, 162, 220, \dodoi{10.3847/1538-3881/ac1c6a}

\bibitem[{Tamayo {et~al.}(2016)Tamayo, Silburt, Valencia, Menou, Ali-Dib, Petrovich, Huang, Rein, Van~Laerhoven, Paradise, {et~al.}}]{Tamayo16}
Tamayo, D., Silburt, A., Valencia, D., {et~al.} 2016, The Astrophysical Journal Letters, 832, L22

\bibitem[{Tamayo {et~al.}(2020)Tamayo, Cranmer, Hadden, Rein, Battaglia, Obertas, Armitage, Ho, Spergel, Gilbertson, {et~al.}}]{Tamayo20}
Tamayo, D., Cranmer, M., Hadden, S., {et~al.} 2020, Proceedings of the National Academy of Sciences, 117, 18194

\bibitem[{Tremaine(2015)}]{Tremaine15}
Tremaine, S. 2015, The Astrophysical Journal, 807, 157

\bibitem[{Van~Eylen \& Albrecht(2015)}]{vanEylen15}
Van~Eylen, V., \& Albrecht, S. 2015, The Astrophysical Journal, 808, 126

\bibitem[{Volk \& Gladman(2015)}]{Volk15}
Volk, K., \& Gladman, B. 2015, The Astrophysical Journal Letters, 806, L26

\bibitem[{{Weiss} {et~al.}(2023){Weiss}, {Millholland}, {Petigura}, {Adams}, {Batygin}, {Block}, \& {Mordasini}}]{Weiss23}
{Weiss}, L.~M., {Millholland}, S.~C., {Petigura}, E.~A., {et~al.} 2023, in Astronomical Society of the Pacific Conference Series, Vol. 534, Protostars and Planets VII, ed. S.~{Inutsuka}, Y.~{Aikawa}, T.~{Muto}, K.~{Tomida}, \& M.~{Tamura}, 863, \dodoi{10.48550/arXiv.2203.10076}

\bibitem[{Weiss {et~al.}(2018)Weiss, Marcy, Petigura, Fulton, Howard, Winn, Isaacson, Morton, Hirsch, Sinukoff, {et~al.}}]{Weiss18}
Weiss, L.~M., Marcy, G.~W., Petigura, E.~A., {et~al.} 2018, The Astronomical Journal, 155, 48

\bibitem[{Wisdom(1980)}]{Wisdom80}
Wisdom, J. 1980, Astronomical Journal, vol. 85, Aug. 1980, p. 1122-1133., 85, 1122

\bibitem[{Wu {et~al.}(2024)Wu, Malhotra, \& Lithwick}]{Wu24}
Wu, Y., Malhotra, R., \& Lithwick, Y. 2024, arXiv preprint arXiv:2405.08893

\bibitem[{Yang \& Tamayo(2024)}]{Yang24}
Yang, Q., \& Tamayo, D. 2024, The Astrophysical Journal, 968, 20

\bibitem[{Zhu {et~al.}(2018)Zhu, Petrovich, Wu, Dong, \& Xie}]{Zhu18}
Zhu, W., Petrovich, C., Wu, Y., Dong, S., \& Xie, J. 2018, The Astrophysical Journal, 860, 101

\end{thebibliography}
\bibliographystyle{aasjournal}
\end{document}